# Structure and magnetic properties of the Ho$_2$Ge$_2$O$_7$ pyrogermanate

E. Morosan,[1] J. A. Fleitman,[1] Q. Huang,[2] J. W. Lynn,[2] Y. Chen,[2,3] X. Ke,[4] M. L. Dahlberg,[4] P. Schiffer,[4] C. R. Craley[1] and R. J. Cava[1]

[1]Department of Chemistry, Princeton University, Princeton, NJ 08544, USA

[2]NIST Center for Neutron Research, NIST, Gaithersburg, MD 20899, USA

[3] Department of Materials Science and Engineering, University of Maryland, College Park, MD 20742, USA

[4]Department of Physics and Materials Research Institute, Pennsylvania State University, University Park, PA 16802, USA

**Abstract**

We report the anisotropic magnetic properties of Ho$_2$Ge$_2$O$_7$ determined from dc and ac magnetization, specific heat and powder neutron diffraction experiments. The magnetic lanthanide sublattice, seen in our refinement of the tetragonal pyrogermanate crystal structure, is a right-handed spiral of edge-sharing and corner-sharing triangles; the local Ho-O coordination indicates that the crystal field is anisotropic. Susceptibility and magnetization data indeed show that the magnetism is highly anisotropic, and the magnetic structure has the Ho moments confined to the plane perpendicular to the structural spiral. The ordered moment of Ho$^{3+}$, as determined from refinement of the neutron diffraction data, is 9.0 $\mu_B$. Magnetic ordering occurs around 1.6 K. Temperature and field dependent ac susceptibility measurements show that this compound displays spin relaxation phenomena analogous to what is seen in the spin ice pyrochlore system Ho$_2$Ti$_2$O$_7$.

**Introduction**

$Ho_2Ge_2O_7$ is a member of the rare earth pyrogermanate series $R_2Ge_2O_7$, for which complex structural and physical properties have been previously reported. Crystallographic analyses have revealed a low symmetry structure (space group P-1) for light rare earth $R_2Ge_2O_7$ (R = La – Sm, except Pr) variants, a result of the large rare earth radii compared to the smaller rare earths (R = Tb – Lu), which form a tetragonal structure, space group $P4_12_12$ [1]. In this latter structure, which has been previously characterized for the cases of R = Eu, Tb, Er, Tm and Lu, but not Ho, there are four formula units per unit cell, with one unique crystallographic rare earth site. The coordination polyhedron of the rare earth is a distorted pentagonal bipyramid [1-2] with its axis almost parallel to the crystallographic *c* direction. The optical properties [3-4] and magnetic susceptibility of the $R_2Ge_2O_7$ compounds [3,5-8] have often been interpreted in light of the four equivalent magnetic sites per unit cell and their local 5-fold symmetry, though the findings of a Mossbauer study on $Er_2Ge_2O_7$ [9] indicate that the pentagonal crystal field model around the R atoms is invalid at least in that case.

The analogy in crystal-chemical character and local structure of the $R_2Ge_2O_7$ pyrogermanates to the $R_2Ti_2O_7$ pyrochlores, which have been of great interest due to their geometrically frustrated spin-spin interactions, indicates that it is desirable to perform detailed studies of the structural and magnetic properties of the former family. This paper focuses on the R = Ho member of the pyrogermanates, for which the rare earths are arranged in a pattern of edge- and corner-sharing triangles, a different configuration than in the pyrochlores. Our current study of $Ho_2Ge_2O_7$ aims to explain the consequences of this arrangement of the magnetic ions, and, as we show below, the magnetic properties are very different than those of the pyrochlores. $Ho_2Ge_2O_7$ crystallizes as tetragonal, transparent plates and is therefore insulating. Although the anisotropic magnetic dc susceptibility has been previously reported [6], no field-dependent measurements exist to date. In addition, the spin freezing that has been observed in the isotypical systems $R_2Ti_2O_7$ for R = Dy [10-11] or Ho [12-13], in which the Ising character of Ho and Dy and the crystal field are critical to the observed spin relaxations, motivated us to perform a study of the frequency- and field-dependent anisotropic ac susceptibility in $Ho_2Ge_2O_7$. In the present study we confirm and further characterize the highly anisotropic magnetic properties. We find that the magnetic moments are confined close to the crystallographic *ab*-plane. Our magnetic neutron diffraction experiments indicate that below a magnetic ordering temperature of 1.6 K, the magnetic unit cell is comprised of eight formula units, with the magnetic spins ordered in a spiral arrangement along the four-fold symmetry axis of the tetragonal cell. We find no evidence of strong magnetic frustration in the magnetization, but the ac susceptibility reveals two possible spin relaxation mechanisms, similar to what is observed in $Ho_2Ti_2O_7$ [12].

**Experimental**

We grew single crystals of $Ho_2Ge_2O_7$ in a flux of $MoO_3$, KF, and $K_2CO_3$, as previously described [14]. The crystals were well-formed, transparent tetragonal plates with typical dimension 3x3x0.5 $mm^3$. We employed powder X-ray diffraction measurements in order to confirm the crystal structure and the purity of the crystals. For neutron diffraction, we made 5g of $Ho_2Ge_2O_7$ by mixing 3.2184 g $Ho_2O_3$ (99.9% Alfa Aesar) and 1.7816 g $GeO_2$ (batch A75054 Johnson Matthey). We first ground the oxides and than heated them in air at $1000^oC$ for 8 hours. Next we reground, pelleted, and heated the mixture in air at $1250^oC$ for 24 hours. We performed neutron diffraction measurements at the NIST Center for Neutron Research, on the high resolution powder neutron diffractometer (BT1) with monochromatic neutrons of wavelength 1.5403 Å and 2.0785 Å produced by Cu(311) and Ge(311) monochromators, respectively. Collimators with horizontal divergences of 15', 20' and 7' of arc were used before and after the monochromator and after the sample, respectively. We collected data in the 2θ range of 3º to 168º with a step size of 0.05°, for a variety of temperatures from 300 K to 5 K to elucidate the magnetic and possible crystal structure transitions. In order to study the magnetic transition properties, we obtained additional diffraction data and magnetic order parameters on the high-intensity triple axis spectrometers at BT7. For these measurements, we used a pyrolytic graphite PG (002) monochromator and filter at a wavelength of 2.36 Å. All structural refinements of the nuclear and magnetic structures in this system were carried out using neutron powder diffraction data and the program GSAS [15]. The neutron scattering amplitudes used in the refinements are 0.808, 0.819, and 0.581 ($cm^{-12}$) for Ho, Ge, and O, respectively.

We performed DC magnetization measurements in a Quantum Design MPMS SQUID magnetometer, and AC susceptibility $\chi_{ac}$ measurements (T = 1.8 - 280 K, H = 5 T, f ≤ $10^4$ Hz) and specific heat data were collected in a Quantum Design PPMS. The specific heat sample was a mixture of $Ho_2Ge_2O_7$ powder with Ag pressed into a pellet. We measured the heat capacity of Ag and sample holder separately, and then subtracted this from the total measured specific heat.

**Results**

The heavy rare-earth pyrogermanates $R_2Ge_2O_7$ (R = Gd - Lu) are reported to be isostructural with $Er_2Ge_2O_7$, which has tetragonal, noncentrosymmetric structure, space group $P4_12_12$ [1]. The magnetic rare-earth sublattice is shown in Fig. 1a. It can be described as a pattern of right-handed helices centered on four fold screw axes parallel to the *c*-axis. Within the helices, the rare earth atoms are locally connected in two edge-sharing triangles. These edge-sharing triangular plaquettes are then joined to each other by corner-sharing triangles. The magnetic lattice can then be considered a spiral of alternating edge-sharing and corner-sharing triangles. The rare-earth ions are coordinated to seven oxygen atoms, forming a distorted pentagonal bipyramid (Fig. 1b). There are five oxygens in a highly distorted pentagon around in the *ab* plane, and oxygens above and below. The distortion of the R-O coordination polyhedron renders an intuitive interpretation of the expected crystal field difficult. The germanium atoms are coordinated to four oxygens in a distorted tetrahedron (Fig. 1c).

Fig. 2a displays the temperature dependent anisotropic magnetization measurements for $Ho_2Ge_2O_7$ for field H || *ab* (circles) and H || *c* (triangles). We observed no indication of magnetic ordering down to the lowest measured temperature T = 1.8 K (left inset, Fig. 2a). The magnetization values are much larger for field applied parallel to the *ab*-plane (circles, left inset Fig. 2a) than for H || *c* (triangles). This indicates that the spins lie close to the *ab*-plane. The anisotropic inverse susceptibility curves (symbols, Fig. 2a) are linear at high temperatures with the H || *c* data displaying a broad peak around 60-80 K (right inset, Fig.2a), possibly associated with the crystal electric field (CEF) anisotropy [6]. At lower temperatures, the "hard axis" (H || c) susceptibility increases as if approaching a low-temperature magnetic ordering. Indeed the specific heat and neutron diffraction data discussed below confirm the presence of a transition temperature below 1.6 K, and this yields a local minimum observed in the H || c susceptibility in the right inset in Fig. 2a. The effective moment values determined from linear fits of the H || *ab*, H || *c* and average inverse susceptibilities above 120 K are $\mu_{eff}^{experiment} \approx 10.2$ $\mu_B$, 9.9 $\mu_B$ and 10.0 $\mu_B$ respectively, somewhat smaller than the theoretical free-ion $\mu_{eff}^{theory} \approx 10.6$ $\mu_B$ expected for $Ho^{3+}$ ions. The reduced $\mu_{eff}$ values in $Tb_2Ge_2O_7$ [8] and $Ho_2Ge_2O_7$ [7] have been attributed to the

complex CEF level splitting. In our measurements the CEF anisotropy is further reflected in the Weiss temperature values $\theta_{ab}$ = 3.8 K (H || ab) and $\theta_c$ = - 65.5 K (H || c), which suggest weak ferromagnetic coupling of the magnetic moment in the *ab*-plane, and much stronger antiferromagnetic interactions in the *c* direction. The anisotropic magnetic properties (effective moment values and Weiss temperatures) are summarized in Table I.

The T = 2 K M(H) isotherms shown in Fig. 2b support the planar magnetic anisotropy picture: the H || *c* axis magnetization (triangles) is small and linear up to our maximum field H = 5 T, as expected for the hard magnetization direction and field values smaller than the CEF energy. For the field applied within the *ab*-plane (circles) a sharp increase of the magnetization below ~ 1 T is followed by a plateau with finite positive slope, attributable to the magnetization slowly rising towards saturation. However, within our field range the maximum magnetization (~ 4 $\mu_B$) is less than half the Ho$^{3+}$ saturated moment $\mu_{sat}$(Ho$^{3+}$) = 10 $\mu_B$.

Specific heat measurements down to 0.8 K reveal the possible presence of a magnetic phase transition. Figure 3 shows the magnetic contribution to the specific heat for Ho$_2$Ge$_2$O$_7$ at low temperatures, where the phonon contributions to the specific heat are subtracted by fitting the data between *T* = 10 - 20 K to the Debye model. The H = 0 curve (full symbols) has a sharp peak near 1.6 K, characteristic of a three dimensional magnetic ordering transition, preceded by a broader peak just below 3 K, associated with the development of short range magnetic correlations. A relatively small applied field of H = 0.5 T (open symbols) suppresses the 3D magnetic ordering transition, while moving the broad specific heat feature to higher temperatures. This suggests that the short range correlations that develop above the 3D ordering temperature are ferromagnetic in character while the long range ordering has a significant antiferromagnetic component.

Given the complex crystal structure described above and its apparent effect on the physical properties of Ho$_2$Ge$_2$O$_7$, it is desirable to determine the magnetic structure of this compound. To this end, we performed low temperature powder neutron diffraction experiments which confirm that Ho$_2$Ge$_2$O$_7$ is isostructural with Er$_2$Ge$_2$O$_7$. The results of the structural investigations on Ho$_2$Ge$_2$O$_7$ are summarized in Table II, where we present the atomic coordinates. We find all positions to have occupancy within one standard deviation of one, all structural parameters well behaved and the agreement between the structural model and the data excellent. The measured neutron diffraction patterns at T = 4.3 K and 1.36 K are presented in Fig. 4, showing that a phase transition has occurred between these temperatures. At T = 4.3 K (Fig. 4a), we are able to fit all observed peaks with the nuclear model based on the space group $P4_12_12$ and lattice parameters a = 6.8083(8) Å and c = 12.3795(2) Å. As the temperature is lowered to T = 1.36 K (Fig. 4b), however, the nuclear model alone cannot account for all the measured peaks, and the character of the differences between the two measurements indicates that the additional peaks observed at the lower temperature are magnetic in origin. Indeed, when we employ a model for the Ho spin ordering (described below), all peaks are well fit, as can be seen in Fig. 5a. Furthermore, the magnetic model for $Ho_2Ge_2O_7$ can be used to completely describe the difference in the patterns measured at T = 1.36 K and 4.3 K (Fig. 5b), a strong indication that the two temperatures are below and above a magnetic phase transition that has little or no structural component [16]. Upon warming the system from 0.5 K, the recorded neutron scattering intensity drops close to zero around T = 1.6 K (Fig. 6), showing that the magnetic phase transition in this system is just below this temperature, consistent with the observations inform the specific heat (Fig. 3). The neutron powder diffraction refinements yield an ordered moment of 9.0 $\mu_B$ per $Ho^{3+}$, close to the maximum value of 10.0 $\mu_B$ for the unperturbed J = 8 $Ho^{3+}$ ion. Thus the low measured magnetization values M(H) ≤ 4 $\mu_B$ (Fig. 2b) could be a consequence of the crystal electric field anisotropy.

The model of the magnetic structure of $Ho_2Ge_2O_7$ describes the Ho spins as being confined to the *ab* plane, as is schematically represented in Fig. 7. The $Ho_2Ge_2O_7$ unit cell contains eight magnetic Ho ions in symmetrically equivalent sites, lying in four layers parallel to the crystallographic *ab* plane. The spins form symmetric pairs around four orthogonal directions in the *ab* plane such that each pair of spins is rotated by 90 degrees from the pairs in the planes above and below. This is best seen in the *ab*-plane projection of the magnetic moments' configuration shown in Fig. 7b.

A comparison of the ordering temperature $T_{ord}$ = 1.6 K with the average Weiss temperature $\theta_{ave}$ = - 9.6 K gives a ratio $|\theta_{ave}|/T_{ord}$ > 5. Within the above magnetic structure picture, this suggests that the geometric frustration expected to exist in this triangle-based system is partially alleviated in the spiral-moment configuration (Fig. 7). However, as we will show below, the low temperature ac susceptibility data of $Ho_2Ge_2O_7$ display signatures of spin freezing resembling the behavior observed in the spin ice system $Dy_2Ti_2O_7$ [10-11] and $Ho_2Ti_2O_7$ [12].

In order to investigate the spin dynamics at low temperature, we measured the anisotropic zero-field cooled (ZFC) ac susceptibility $\chi_{ac} = \chi' + i\chi''$ for variable frequency and applied field, which is shown in Fig. 8. By contrast to the dc susceptibility, which decreased monotonically with temperature above 1.8 K (Fig. 2a), $\chi'$ shows up to two relaxation temperatures marked by broad peaks. One such peak develops below 3 K at the lowest applied field H = 0.1 T (Fig. 8a), and is frequency independent. However, a different response of $\chi'$ is registered as the applied magnetic field increases: in the low frequency limit (f = 10 Hz, Fig. 8b) this peak moves up as it broadens, centering around 16 K at H = 1.2 T. With increasing frequency (f ≥ $10^2$ Hz, Fig. 8c-e) a second lower temperature peak emerges, with much weaker field dependence. The imaginary part $\chi''$ of the ac susceptibility reveals striking similarities with the relaxation behavior observed in the $Ho_2Ti_2O_7$ spin ice [12]: as can be seen in Fig. 8f for H = 0.4 T, $\chi''$ displays a peak just below 10 K, which moves up in temperature as frequency increases. The inset shows the Arrhenius plot for the frequency shift of $Ho_2Ge_2O_7$, indicating that the high-temperature process is likely due to a thermally activated single-ion relaxation process, as has been demonstrated by ac susceptibility and neutron spin echo measurements on $Ho_2Ti_2O_7$ [17]. From the Arrhenius plot (inset Fig.8f) we determined the corresponding activation energy $E_a$ to be around 90 K, comparable with the 60-80 K broad peak seen in the H||c dc susceptibility (right inset, Fig. 2a), and thus $E_a$ is of the order of the CEF energy. In a similar manner to $Ho_2Ti_2O_7$, the frequency of the low-temperature peak observed in $\chi'$ for $Ho_2Ge_2O_7$ (Fig. 8b-e) has a weak field and temperature dependence. The other field orientation (H||c) yields a field- and frequency-independent $\chi'$, and a representative field manifold is shown in Fig. 9 for f = $10^3$ Hz.

Although the frequency-dependent freezing could be indicative of spin glass behavior, the field and frequency dependence of the ac susceptibility in $Ho_2Ge_2O_7$ is qualitatively comparable to that of $Ho_2Ti_2O_7$ spin ice [12,13,17] in that the applied field enhances the higher relaxation temperature $T_1$ (full symbols, Fig. 10) yielding an order of magnitude increase of $T_1$ between 0 and 1.2 T. The lower relaxation temperature $T_2$ is virtually unaffected by the applied field and is also almost frequency independent (open symbols, Fig. 10).

**Discussion**

A new spin freezing mechanism called spin ice freezing was reported for the pyrochlore ferromagnets $Dy_2Ti_2O_7$, $Ho_2Ti_2O_7$ and, more recently, $Ho_2Sn_2O_7$ [10-11,18]. This freezing is associated with the development of local correlations among neighboring spins as they dynamically freeze into a disordered low temperature state associated with the frustration of spin-spin interactions. Similar to spin-glass-like freezing, the ac susceptibility of spin-ice has a frequency dependent local maximum, but the two mechanisms are set apart by the field dependence of the freezing temperature $T_f$ and the frequency distribution of the relaxation times. $Ho_2Ge_2O_7$ shows long range order at low magnetic fields, but it displays behavior similar to spin-ice freezing in a sufficiently strong magnetic field: up to two local maxima exist in the ac susceptibility, with one transition being driven up by the application of magnetic field, and the second one almost field and frequency independent. For $Ho_2Ge_2O_7$ the frequency dependence of the low-temperature relaxation process can be fit to the Arrhenius relation $f = f_0 \exp(-E_a/kT_f)$, with $E_a \approx 90$ K, close to the CEF energy inferred from the dc susceptibility. Further studies are needed to clarify whether the behavior observed in $Ho_2Ge_2O_7$ and $Ho_2Ti_2O_7$ indeed results from similar development of local correlations; in particular, measurements on a non-magnetic dilution of $Ho_2Ge_2O_7$ might confirm the high temperature single-ion scenario, and magnetization measurements below 1.8 K are necessary to clarify the nature of the low-temperature transition. Additionally, neutron diffraction experiments in applied magnetic field could further elucidate the nature of the magnetic structure in this compound, and this is left to a future study.

**Acknowledgements**

The work at Princeton was supported by grants NSF DMR02-13706, and DOE-FG-02-45706. The work at Penn State was supported by NSF grant DMR-0701582. Identification of commercial equipment in the text is not intended to imply recommendation or endorsement by the National Institute of Standards and Technology.

**References**

1. U. W. Becker and J. Felsche, *J. Less-Common Metals* **128** (1987) 269.
2. Yu. I. Smolin , *Sov. Phys. Crystallogr.* **15** (1970) 36.
3. K. Das, S. Jana, D. Ghosh and B. M. Wanklyn, *J. Magn. Magn. Mat.* **189** (1998) 310.
4. D. M. Moran, F. S. Richardson, M. Koralewski and B. M. Wanklyn, *J. Alloys and Comp.* **180** (1992) 171.
5. A. Sengupta, D. Ghosh and B. M. Wanklyn, *Phys. Rev. B* **47** (1993) 8281.
6. S. Jana, D. Ghosh and B. M. Wanklyn, *J. Magn. Magn. Mat.* **183** (1998) 135.
7. M. Ghosh, S. Jana, D. Ghosh and B. M. Wanklyn, *Solid State Commun.* **107** (1998) 113.
8. Y. M. Jana, M. Ghosh, D. Ghosh and B. M. Wanklyn, *J. Magn. Magn. Mat.* **210** (2000) 93.
9. J. M. Cadogan, D. H. Ryan, G. A. Stewart and R. Gagnon, *J. Magn. Magn. Mat.* **265** (2003) 199.
10. J. Snyder, J. S. Slusky, R. J. Cava and P.Schiffer, *Nature* **413** (2001) 48.
11. J. Snyder, B. G. Ueland, J. S. Slusky, H. Karunadasa, R. J. Cava and P.Schiffer, *Phys. Rev. B* **69** (2004) 64414.
12. G. Ehlers, A. L. Cornelius, T. Fennell, M. Koza, S. T. Bramwell and J. S. Gardner, *J. Phys: Cond. Matter* **16** (2004) S635.
13. G. Ehlers, J. S. Gardner, C. H. Booth, M. Daniel, K. C. Kam, A. K. Cheetham, D. Antonio, H. E. Brooks, A. L. Cornelius, S. T. Bramwell, J. Lago, W. Häussler and N. Rosov, *Phys. Rev. B* **73** (2006) 174429.
14. C. R Craley, *Crystal Growth of Some Geometrically Frustrated Magnets*, in *Chemistry*. 2004, Princeton University: Princeton.
15. A.C. Larson and R.B. Von Dreele, Los Alamos National Laboratory Report No. LAUR086-748 (1990).
16. For details of the subtraction technique see H. Zhang, J. W. Lynn, W-H. Li, T. W. Clinton, and D. E. Morris, Phys. Rev. B **41**, (1990) 11229.
17. G. Ehlers, A. L. Cornelius, M. Orendac, M. Kajnakova, T. Fennell, T. Fennell, S. T. Bramwell and J. S. Gardner, *J. Phys: Cond. Matter* **15** (2003) L9.
18. H. Kadowaki, Y. Ishii, K. Matsuhira, Y. Hinatsu Phys. Rev. B **65**, (20020) 144421.

Table I. Anisotropic magnetic properties for $Ho_2Ge_2O_7$: The effective moment values $\mu_{eff}$ and the Weiss temperatures $\theta_W$ are determined from linear fits of the inverse susceptibility at temperatures above 120 K.

|  | H ∥ *ab* | H ∥ *c* | average | theory |
|---|---|---|---|---|
| $\mu_{eff}(\mu_B)$ | 10.2 | 9.9 | 10.0 | 10.6 |
| $\theta_W$ (K) | 3.8 | - 65.5 | - 9.6 |  |

Table II. Refined atomic positional coordinates of $Ho_2Ge_2O_7$, space group $P4_12_12$ (92), unit cell parameters a = 6.8083(8) Å, c= 12.3795(2) Å. Residuals for the fit: $\chi^2$ = 1.02, $wR_p$ = 2.6%, $R_p$ = 2.1%.

| Site | Wyckoff Symbol | x | y | z | Occ | Beq |
|---|---|---|---|---|---|---|
| Ho | 8a | 0.87664(11) | 0.35295(13) | 0.13499(8) | 1.00 | 0.80(2) |
| Ge | 8a | 0.90095(12) | 0.15382(14) | 0.61957(8) | 1.00 | 1.06(2) |
| O1 | 4a | 0.80458(20) | 0.19542(20) | 0.750000 | 1.00 | 1.50(4) |
| O2 | 8a | 0.07707(20) | -0.03060(19) | 0.62405(13) | 1.00 | 1.33(3) |
| O3 | 8a | 0.06207(19) | 0.33815(22) | 0.57183(11) | 1.00 | 1.30(3) |
| O4 | 8a | 0.68611(17) | 0.14240(22) | 0.54501(10) | 1.00 | 1.41(3) |

Fig. 1. Crystal structure of Ho$_2$Ge$_2$O$_7$: (a) the rare earth R sublattice showing the right-handed helices centered on four-fold screw axes parallel to the c-axis; the edge-sharing triangles connecting the Ho atoms are shaded. Different colored atoms illustrate the four R atoms around the four-fold screw axis. (b) the R – O and (c) R - Ge coordination polyhedra, with R – red (dark grey online) and O/Ge – white/light blue (light grey online): there are five O in a highly distorted pentagon in the ab-plane, and two more O, one above and one below.

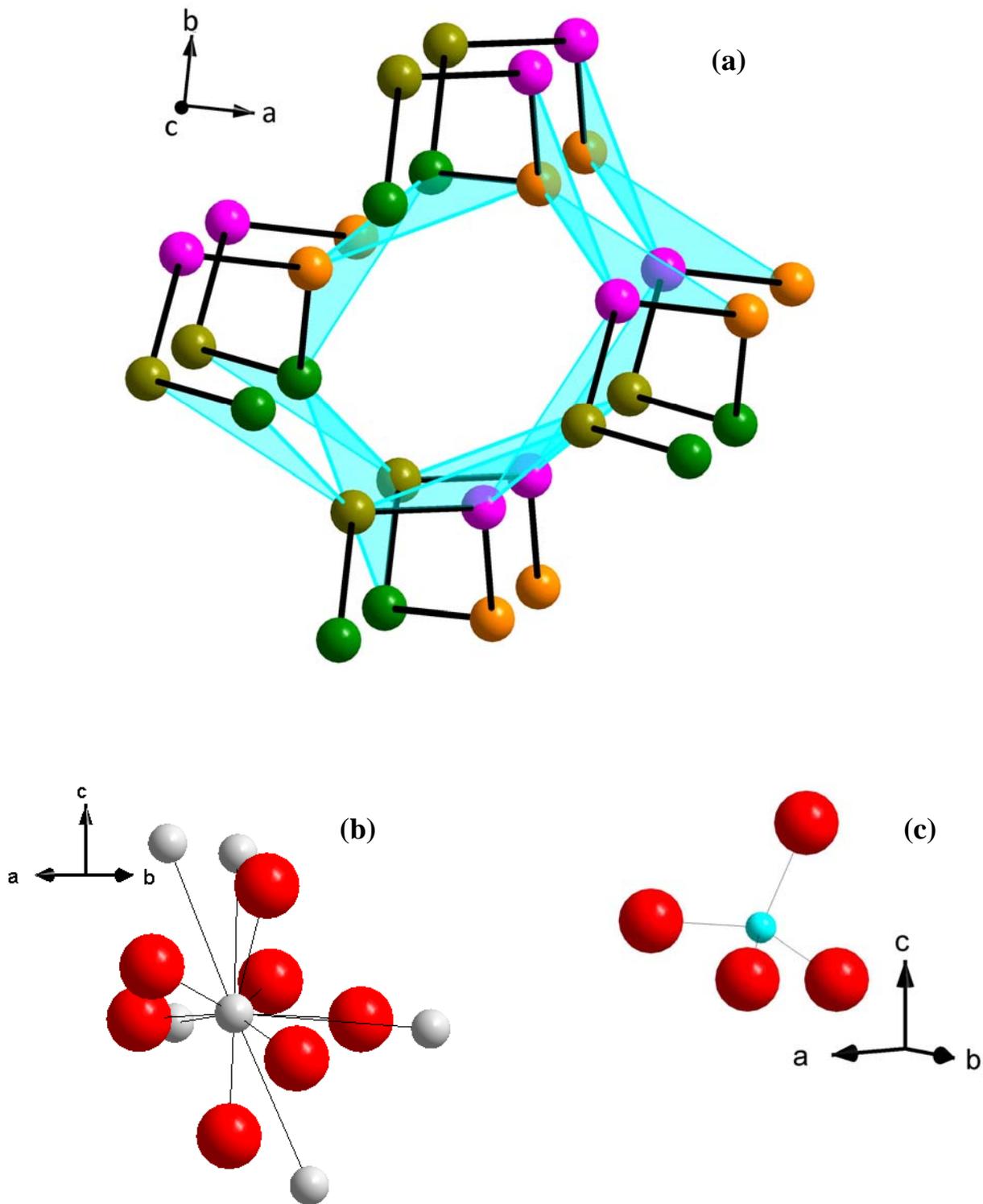

Fig. 2. (a) Inverse magnetic susceptibility for H = 0.1 T and H||ab (circles) and H||c (triangles) together with the calculated average inverse susceptibility (line). Inset: low temperature anisotropic susceptibilities. (b) T = 2 K M(H) isotherms for H||ab (circles) and H||c (triangles).

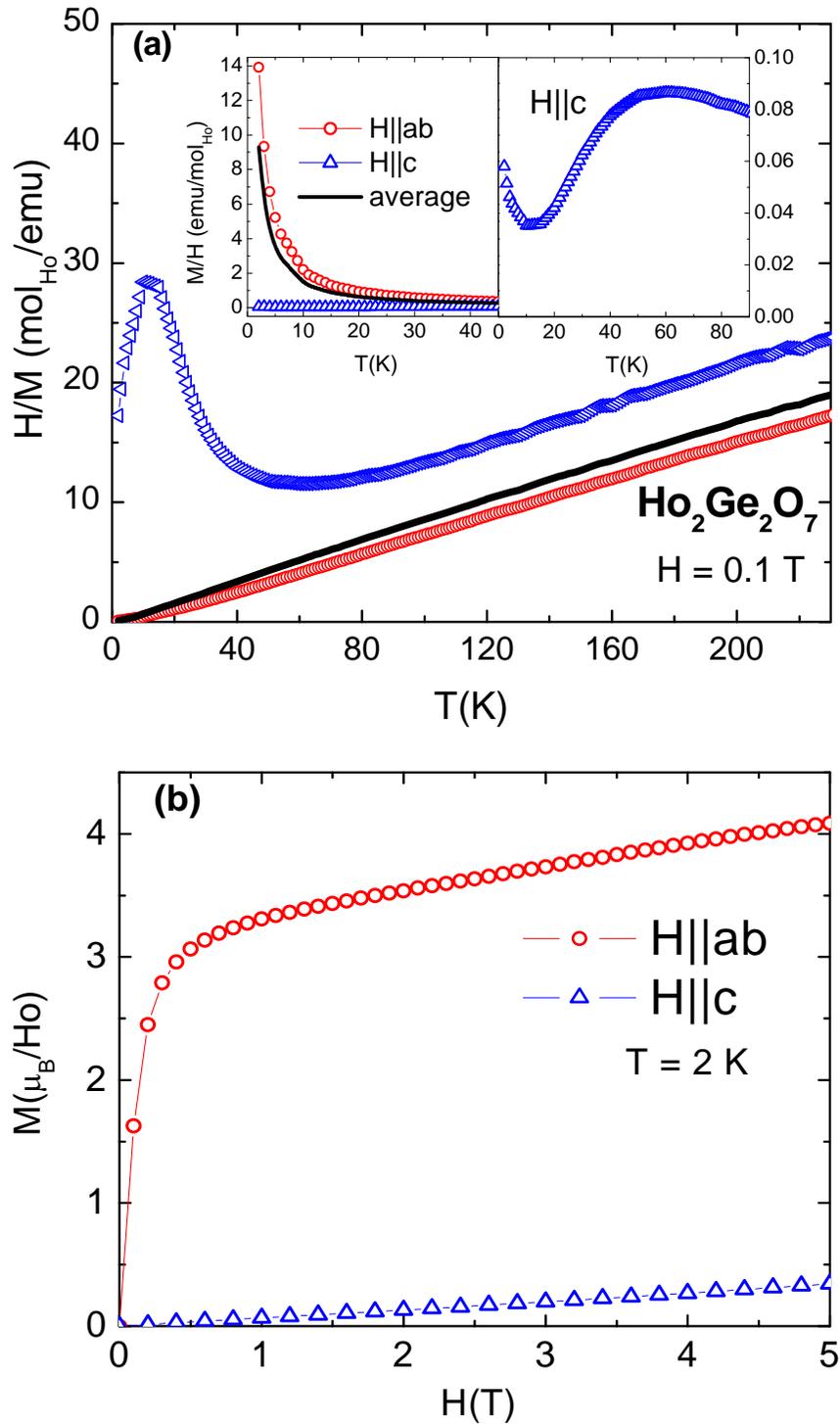

Fig. 3. Specific heat for $Ho_2Ge_2O_7$ measured in zero (full symbols) and applied field H = 0.5 T (open symbols).

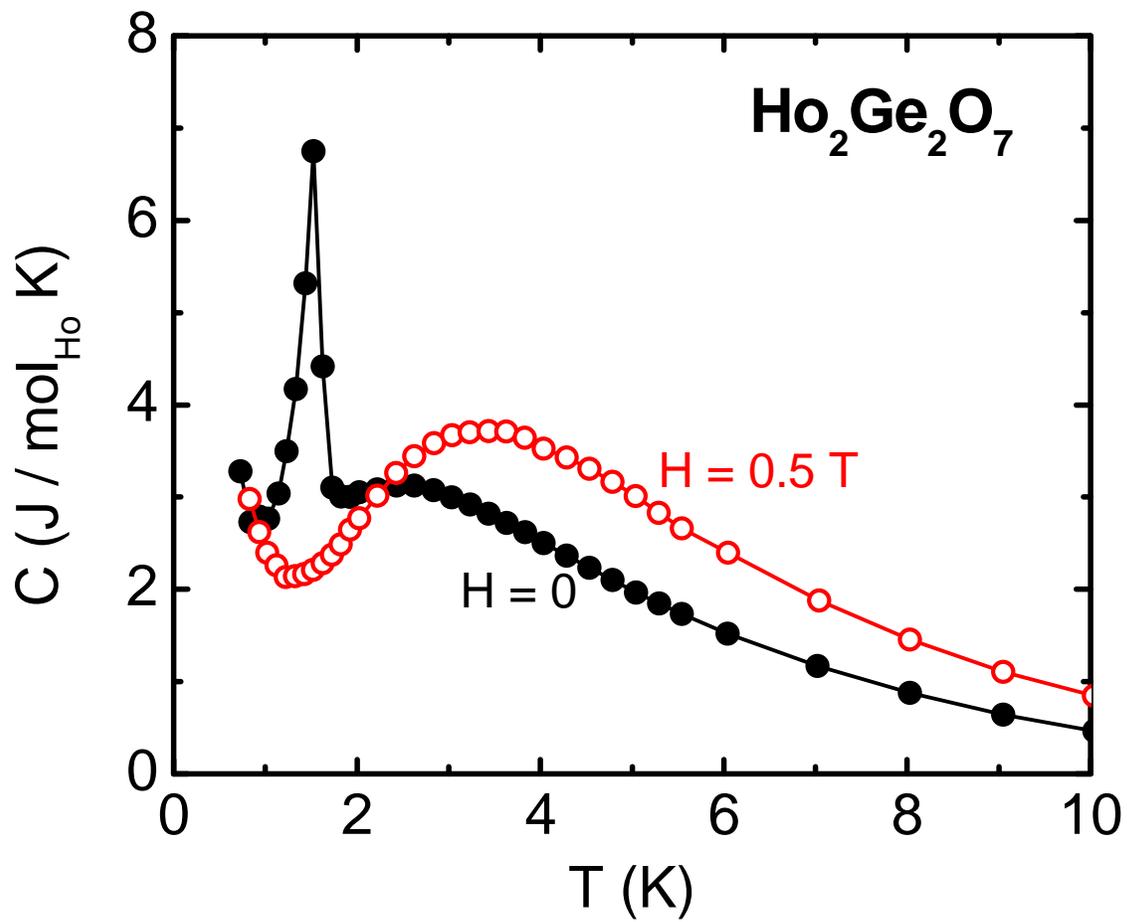

Fig. 4. Neutron diffraction measurements (red) at (a) T = 4.3 K and (b) T = 1.36 K together with a fit using the nuclear model only (see text) and the difference between measured and model calculation (purple). In (b) the crystallographic reflections have been removed from the top pattern to show just the magnetic intensities in the bottom pattern.

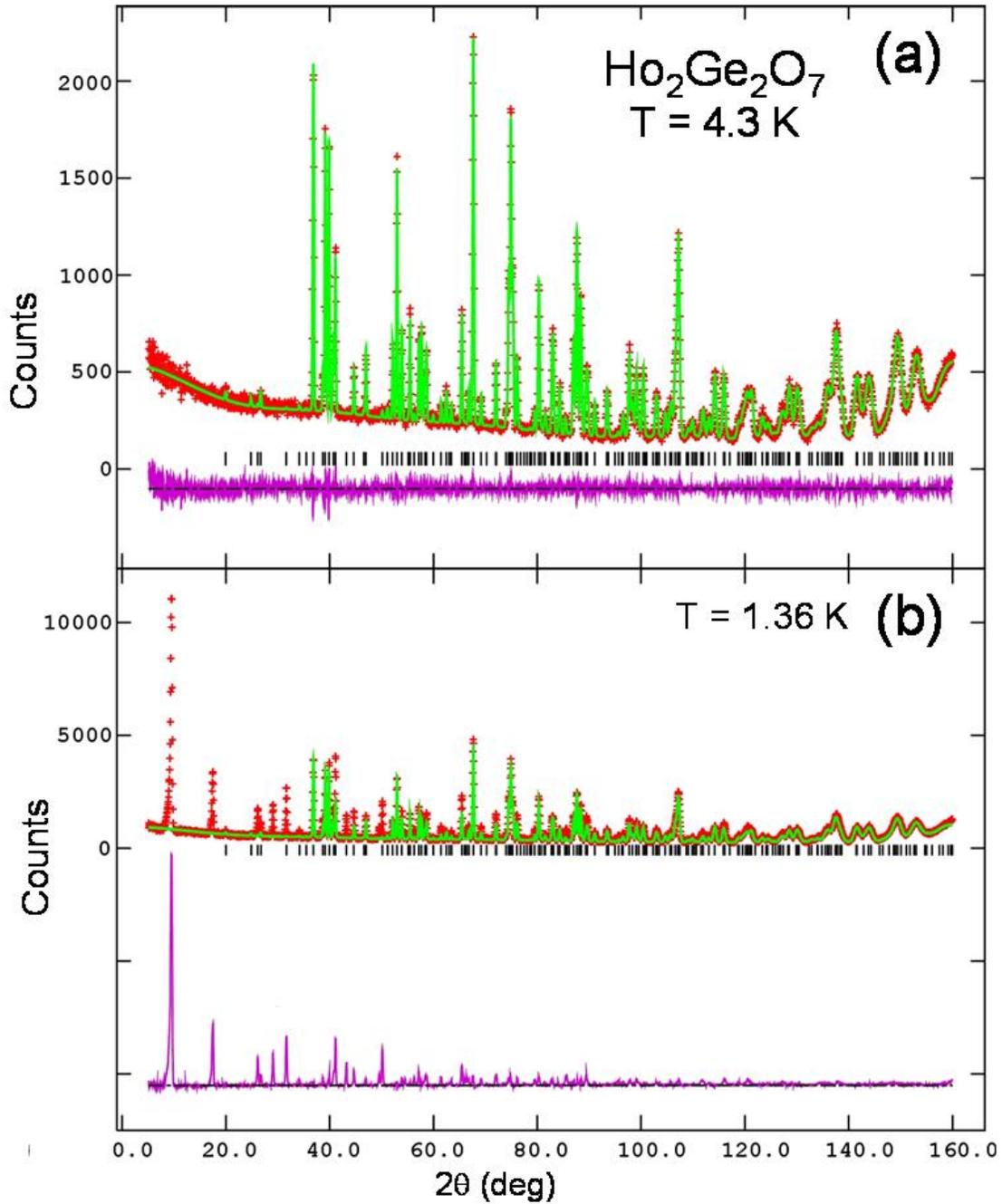

Fig. 5. Difference between T = 1.36 K and T = 4.3 K neutron diffraction measurements (red) together with a fit using (a) the magnetic model only (see text) and (b) the nuclear and magnetic model; purple curve: difference between experimental and model patterns. The magnetic structure model used in the refinements has the magnetic space group symmetry $P4_12_12$ and the refined magnetic moment per Ho at 0.3 K is 9.05(4) $\mu_B$, with components $\mu_x$ =1.67(4) $\mu_B$, $\mu_y$ =-8.90(4) $\mu_B$ and $\mu_z$ =0 $\mu_B$.

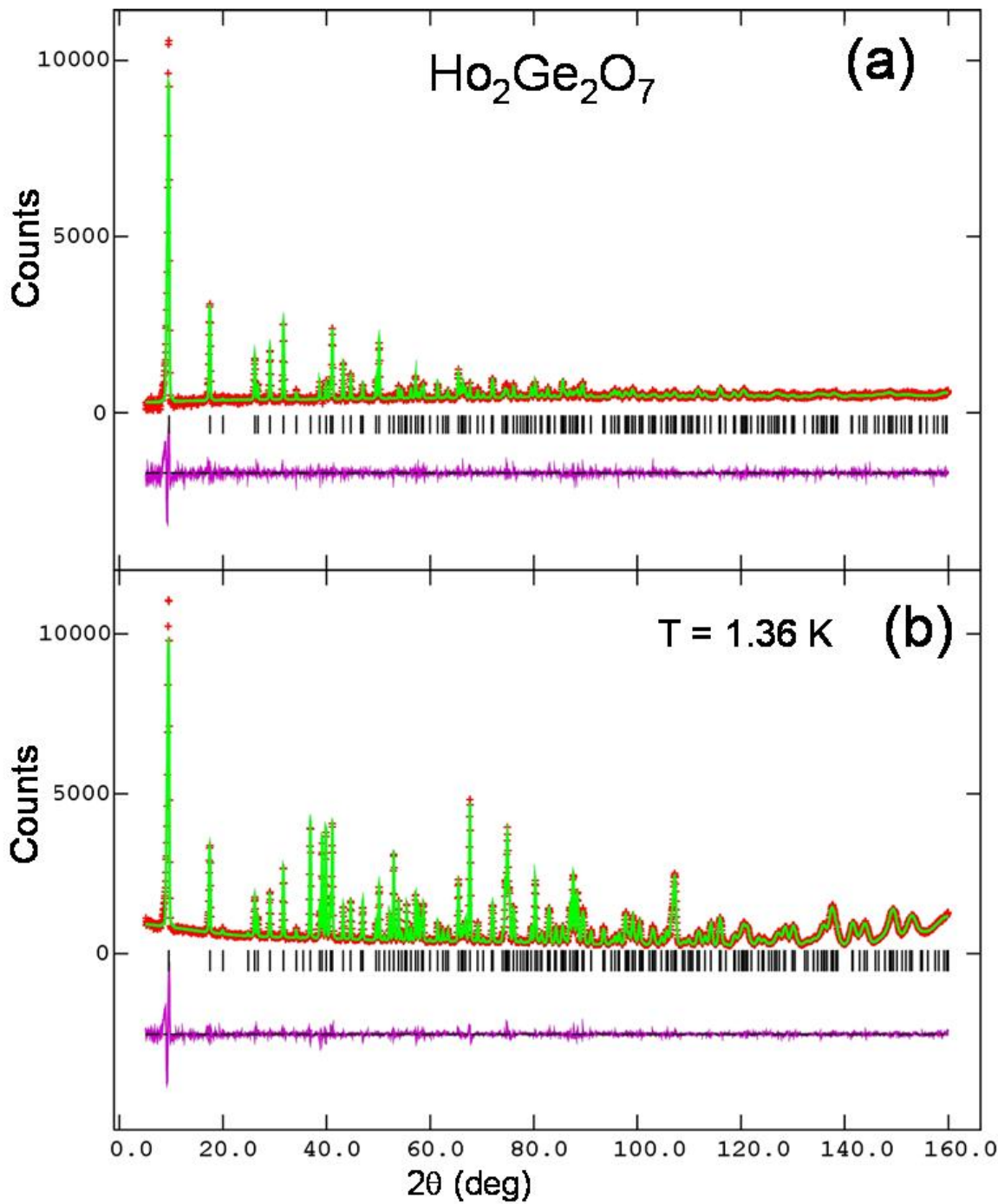

Fig. 6. Neutron diffraction intensities measured on warming (full symbols) and on cooling (open symbols) around the magnetic phase transition at T ≈ 1.6 K.

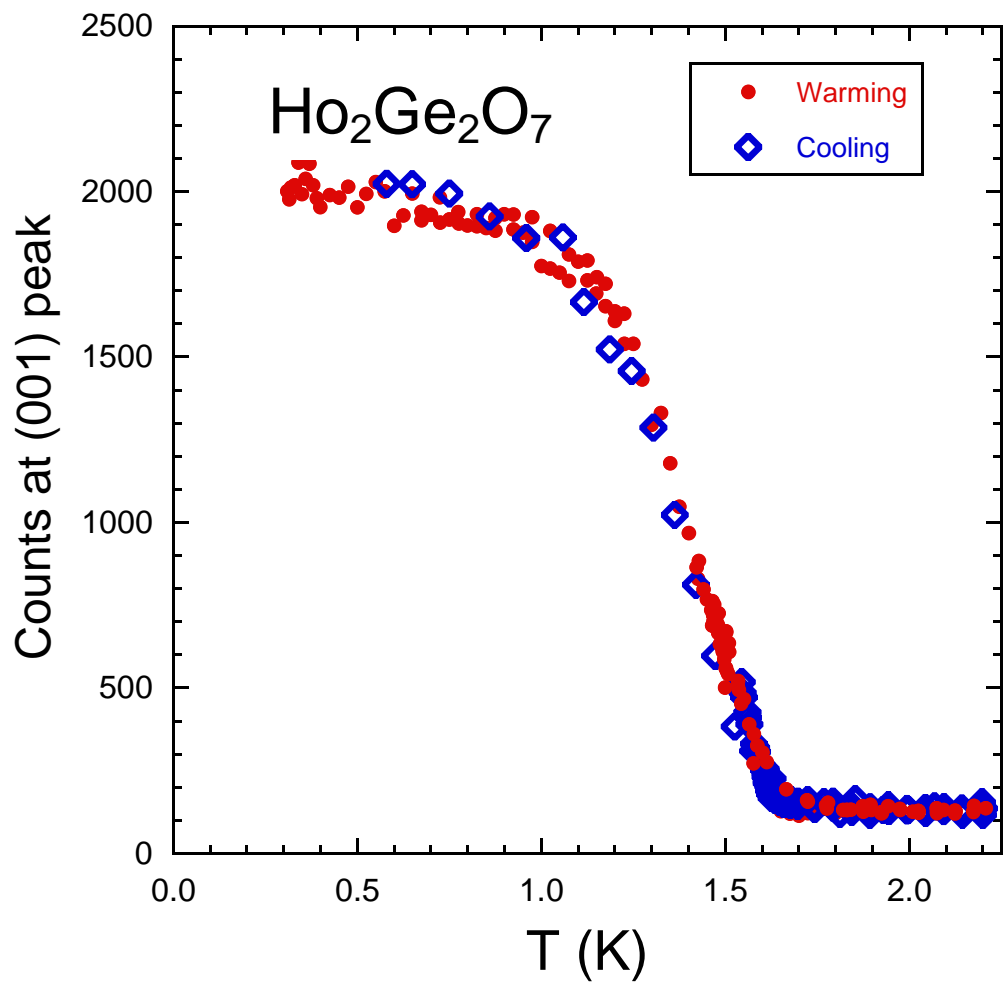

Fig. 7. (a) Magnetic spin configuration showing eight Ho ions in four equivalent ab planes. (b) ab projection of the Ho lattice, showing that pairs in subsequent planes are rotated by 90° from those in the planes above and below.

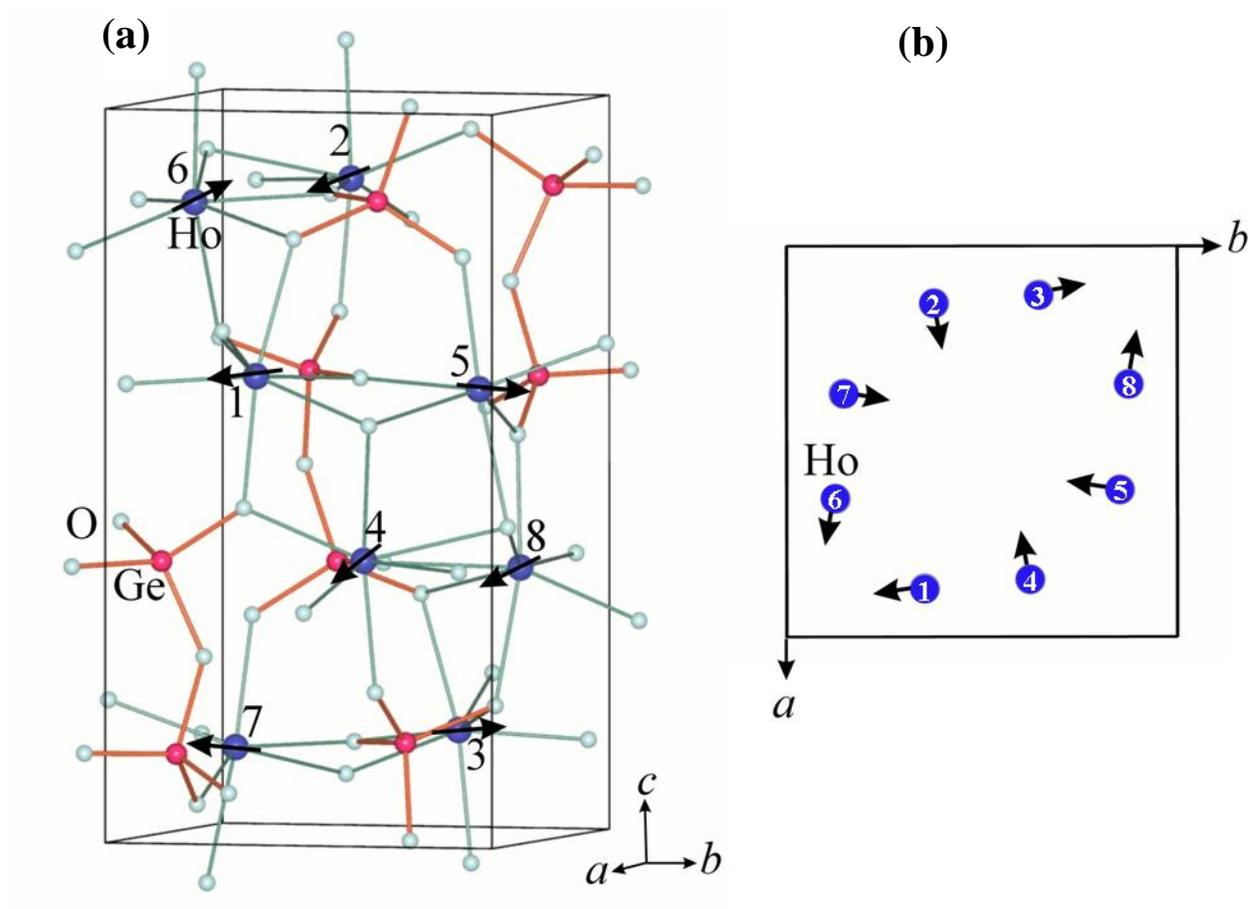

Fig. 8. H∥ab ac susceptibility data: (a) $\chi'$ for H = 0.1 T and f = 10, $10^2$, $10^3$ and $10^4$ Hz. (b-e) f = 10, $10^2$, $10^3$ and $10^4$ Hz data for H = 0.1 T, 0.4 T, 0.5 T, 0.6 T, 0.7 T, 0.8 T, 1.0 T and 1.2 T. (f) $\chi''$ for H = 0.4 T and f = 10, $10^2$, $10^3$ and $10^4$ Hz, with the Arrhenius plot log f ($T^{-1}$) shown in the inset.

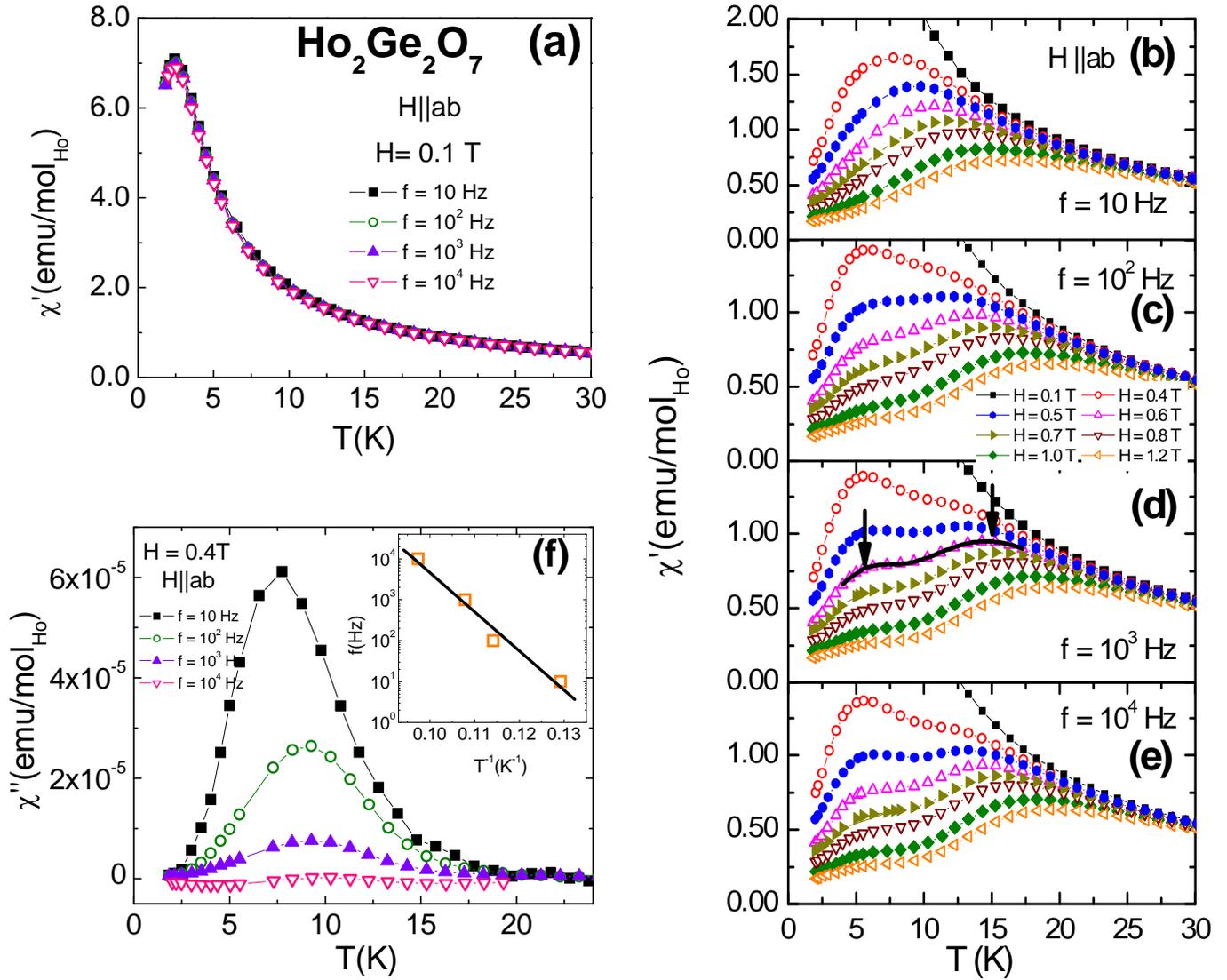

Fig. 9. Real part χ' of the H||c ac susceptibility data for H = 0.1T, 0.5 T, 0.8 T and 1.0 T T and f = $10^3$ Hz.

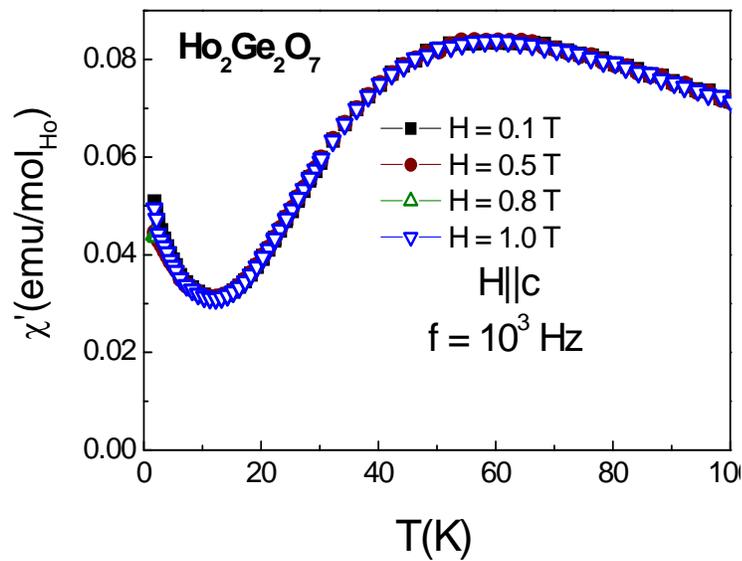

Fig. 10. The two relaxation temperatures $T_1$ and $T_2$ as a function of applied field for H||ab; lines are guides for the eye, showing the increase of $T_1$ with H, and almost field-independent $T_2$.

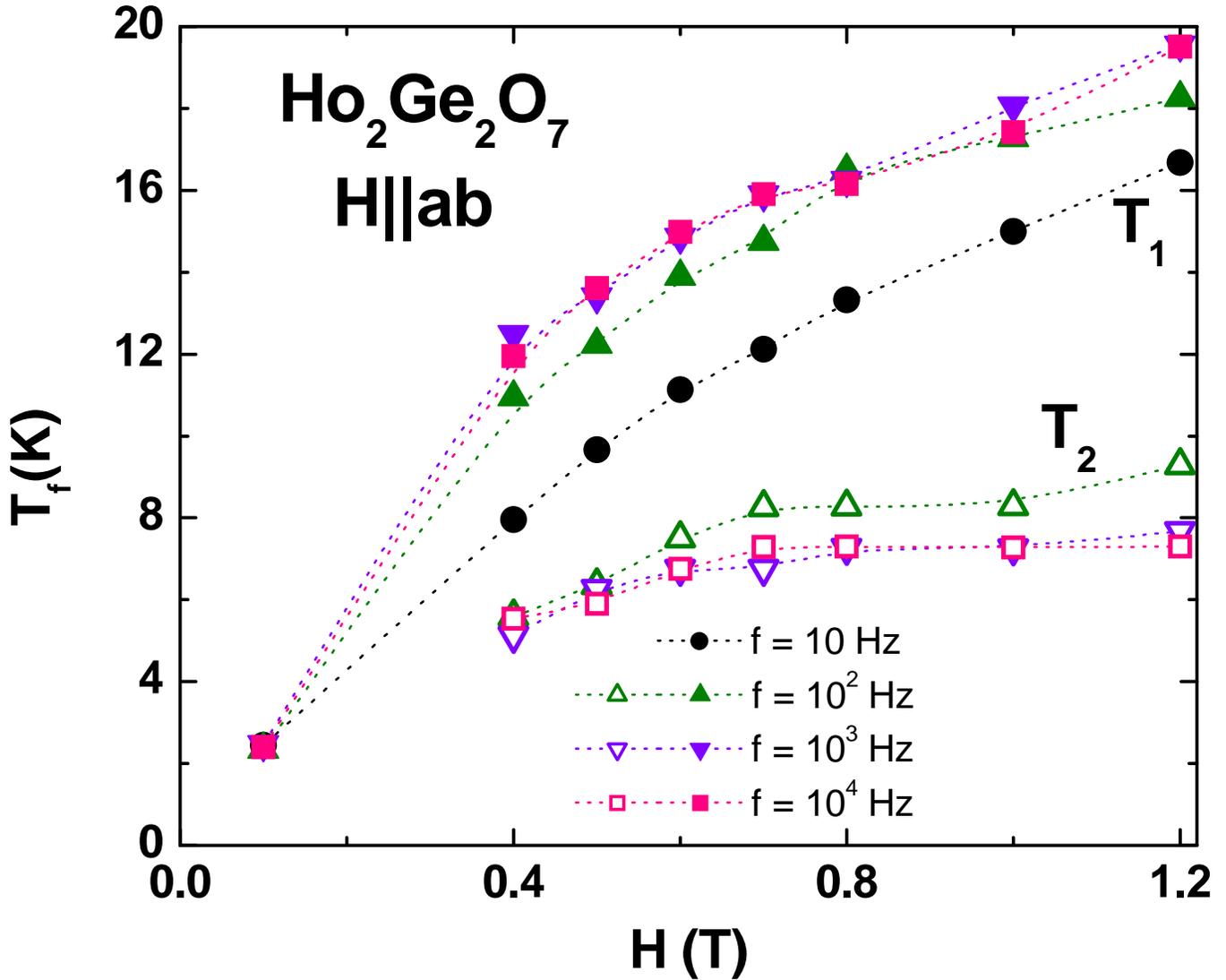